# Robust Adaptive Sparse Channel Estimation in the Presence of Impulsive Noises

Guan Gui, Li Xu
Dept. of Electronics and Information Systems
Akita Prefectural University
Yurihonjo, Japan
Emails: {guiguan,xuli}@akita-pu.ac.jp

Wentao Ma, Badong Chen
School of Electronic and Information Engineering
Xi'an Jiaotong University
Xi'an, China
Email: chenbd@mail.xjtu.edu.cn

*Abstract*—**Broadband wireless channels usually have the sparse nature. Based on the assumption of Gaussian noise model, adaptive filtering algorithms for reconstruction sparse channels were proposed to take advantage of channel sparsity. However, impulsive noises are often existed in many advance broadband communications systems. These conventional algorithms are vulnerable to deteriorate due to interference of impulsive noise. In this paper, sign least mean square algorithm (SLMS) based robust sparse adaptive filtering algorithms are proposed for estimating channels as well as for mitigating impulsive noise. By using different sparsity-inducing penalty functions, i.e., zero-attracting (ZA), reweighted ZA (RZA), reweighted L1-norm (RL1) and Lp-norm (LP), the proposed SLMS algorithms are termed as SLMS-ZA, SLMS-RZA, LSMS-RL1 and SLMS-LP. Simulation results are given to validate the proposed algorithms.**

## I. INTRODUCTION

Broadband transmission is becoming more and more important technique in advanced wireless communications systems [1]–[3]. The main impairments to the systems are due to multipath fading propagation as well as additive noises. Hence, accurate channel state information (CSI) is required for coherence detection [1]. Based on the assumption of Gaussian noise model, channel estimation has been extensively studies in the literatures [4]–[8]. However, these methods may unsuitable to apply directly in estimating channel under non-Gaussian noise environment. For example, the performance of SOS-LMS [4] is vulnerable to deteriorate by strong impulsive noise interference in advanced communications systems [9]. Such impulsive noise, which results from natural or man-made electromagnetic waves, usually has heavy-tailed distribution and violates the commonly used Gaussian noise assumption [10]. Hence, accurate noise model is one of important technical issues for designing dependable systems.

The aforementioned impulsive noise can be described by the family of alpha-stable distributions [11] which can also model many impulsive noise processes in communications channels and in fact, includes the Gaussian density as a special case. To mitigate the harmful interferences, it is necessary to develop robust channel estimation algorithms. Based on the assumption of dense finite impulse response (FIR), recently, several effective adaptive channel estimation algorithms have been proposed to achieve the robustness against impulsive interferences [12]–[14]. In [12], variable step-size (VSS) sign algorithm based adaptive channel estimation was proposed to achieve performance gain. In [13], an standard VSS affine projection sign algorithm (VSS-APSA) was proposed and its improved version was also proposed in [14]. However, FIR of real wireless channel is often modeled as sparse or cluster-sparse and hence many of channel coefficients are zeros [15] [16]. However, these algorithms may unable to exploit sparse channel structure information and accurately there are some performance gain could be obtained if we can develop advanced adaptive channel estimation methods.

In this paper, we propose four sparse SLMS algorithms by adopting four sparsity-inducing penalty functions, i.e., zero-attracting (ZA) [7], reweighted zero attracting (RZA) [7], reweighted $\ell_1$-norm (RL1) [17] and $\ell_p$ -norm (LP), to exploit channel sparsity as well as to mitigate impulsive interferences. Our contribution of this paper can be summarized as follows. First, cost function of SLMS-ZA is constructed and the corresponding update equation is derived. Second, SLMS-RZA, SLMS-RL1 and SLMS-LP are developed as well. At last, several representative simulation results are provided to verify the effectiveness of the proposed algorithms.

## II. SYSTEM MODEL AND PROBLEM FORMULATION

Let us consider an additive alpha-stable noise interference channel, which is modeled by the unknown *N*-length finite impulse response (FIR) vector $\boldsymbol{w} = [w_0, w_1, \cdots, w_{N-1}]^T$ at discrete time-index $n$. The ideal received signal is obtained as

$$d(n) = \boldsymbol{w}^T \boldsymbol{x}(n) + z(n), \qquad (1)$$

where $\boldsymbol{x}(n) = [x(n), x(n-1), \cdots, x(n-N+1)]^T$ is the input signal vector of the $N$ most recent input samples and $z(n)$ is a $\alpha$-stable noise. The characteristic function of alpha-stable process is defined as

$$p(t) = \exp\{j\delta t - \gamma |t|^\alpha [1 + j\beta \operatorname{sgn}(t)\mathcal{S}(t,\alpha)]\}, \quad (2)$$

where

$$S(t,\alpha) = \begin{cases} \tan(\alpha\pi/2), & \text{if } \alpha \neq 1 \\ (-2/\pi)\log(t), & \text{if } \alpha = 1 \end{cases} \quad (3)$$

Here, $\alpha \in (0,2]$ denotes the characteristic exponent to measure the tail heaviness of the distribution, i.e., smaller $\alpha$ indicates heavier tail and vice versa. In addition, one can find that the Gaussian process is a special case of the paper, we assume that alpha-stable noise when $\alpha = 2$. $\gamma > 0$ represents the dispersive parameter to act a similar role to the variance of Gaussian distribution; $\beta \in [-1,1]$ denotes the symmetrical parameter which controls symmetry scenarios about its local parameter $\delta$. Throughout noise is symmetrical in the case of $\beta = 0$ as well as $\delta = 0$. The objective of adaptive channel estimation is to perform adaptive estimate of $w(n)$ with limited complexity and memory given sequential observation $\{d(n), x(n)\}$ in the presence of additive noise $z(n)$. That is to say, the estimate observation signal $y(n)$ is given as

$$y(n) = w^T(n)x(n), \quad (4)$$

where $w(n)$ is an $N$-dimensional signal vector of the unknown system; $z(n)$ describes the measurement noise with variance $\sigma_n^2$. By combining (1) and (4), the estimation error $e(n)$ is

$$e(n) = d(n) - y(n) = z(n) - x^T(n)v(n), \quad (5)$$

where $v(n) = w - w(n)$ is the estimate error of $w(n)$ at iteration $n$. The cost function of standard LMS was written as

$$G_{ZA}(w(n)) = (1/2)e^2(n). \quad (6)$$

The update equation of LMS was derived as

$$w(n+1) = w(n) + \mu \frac{\partial G(w(n))}{\partial w(n)} = w(n) + \mu e(n)x(n), \quad (7)$$

where $\mu$ is step-size. To keep stable of gradient descend, the range of $\mu$ is chosen as $\mu \in (0,2)$. To mitigate impulsive noise, then the standard SLMS [12] was proposed as

$$w(n+1) = w(n) + \mu \text{sgn}(e(n))x(n), \quad (8)$$

where $\text{sgn}(\cdot)$ denotes sign function, i.e., $\text{sgn}(x) = 1$ for $x > 0$, $\text{sgn}(x) = -1$ for $x < 0$ and $\text{sgn}(x) = 0$ for $x = 0$.

## III. STABLE SPARSE ADAPTIVE FILTERING ALGORITHMS

To full take advantage of channel sparsity, optimal sparse constraint function (i.e., $\ell_0$-norm) [18] is utilized as sparse SLMS algorithm for estimating channels in impulsive interference environments. By virtue of the $\ell_0$-norm as for sparsity-inducing penalty function on channel estimate $w(n)$, mathematically, the cost of function of optimal sparse SLMS is constructed as

$$G_{L0}(w(n)) = (1/2)e^2(n) + \lambda_{L0}\|w(n)\|_0, \quad (9)$$

where $\|\cdot\|_0$ represents $\ell_0$-norm function and $\lambda_{L0}$ denotes the regularization parameter to balance the instantaneous updating error and sparsity of $w(n)$. In the perspective of mathematical theory, $\ell_0$-norm can exploit most sparsity information on sparse channel estimation. However, computing the $\ell_0$-norm is a NP-hard (non-deterministic polynomial-time hard) problem [18]. Hence, it is necessary to consider computable sparse constraints on sparse channel estimation. By means of four sparsity functions, in the subsequent, four sparse SLMS algorithms, i.e., SLMS-ZA, SLMS-RZA, SLMS-RL1 and SLMS-LP, are proposed to exploit the channel sparsity as well as to mitigate the impulsive interferences simultaneously.

### A. Proposed algorithm: SLMS-ZA

According to (9), one can replace the $\ell_0$-norm sparse constraint with $\ell_1$-norm function [19]. Then, cost function of LMS-ZA was constructed as

$$G_{ZA}(w(n)) = (1/2)e^2(n) + \lambda_{ZA}\|w(n)\|_1, \quad (10)$$

where $\lambda_{ZA}$ denotes a regularization parameter to balance estimation error and $\ell_1$-norm sparsity function of the $w(n)$. According to (10), hence, the update equation of LMS-ZA was derived as

$$\begin{aligned} w(n+1) &= w(n) - \mu \frac{\partial G_{ZA}(w(n))}{\partial w(n)} \\ &= w(n) + \mu e(n)x(n) - \rho_{ZA}\text{sgn}(w(n)). \end{aligned} \quad (11)$$

where $\rho_{ZA} = \mu\lambda_{ZA}$ is decided by $\mu$ and $\lambda_{ZA}$. With respect to estimate vector $w(n)$, SLMS-ZA is proposed as

$$w(n+1) = w(n) + \mu\text{sgn}(e(n))x(n) - \rho_{ZA}\text{sgn}(w(n)). \quad (12)$$

### B. Proposed algorithm: SLMS-RZA

It was well known that more strong sparse constraint could exploit sparsity more efficient [17]. This principle implies that channel estimation performance could be improved by using more efficient sparse approximation function even if in the presence of impulsive noises. Hence, the cost function of LMS-RZA is written as

$$G_{RZA}(w_i(n)) = \frac{1}{2}e^2(n) + \lambda_{RZA}\sum_{i=0}^{N-1}\log(1+\varepsilon_{RZA}|w_i(n)|), \quad (13)$$

where $\lambda_{RZA} > 0$ is a regularization parameter to balance the estimation error and sparsity of $\sum_{i=0}^{N-1}\log(1+\varepsilon_{RZA}|w_i(n)|)$. Likewise, the corresponding update equation was derived as

$$\begin{aligned} w_i(n+1) &= w_i(n) + \mu \frac{\partial G_{RZA}(w_i(n))}{\partial w_i(n)} \\ &= w_i(n) + \mu e(n)x(n-i) - \frac{\rho_{RZA}\text{sgn}(w_i(n))}{1+\varepsilon_{RZA}|w_i(n)|}. \end{aligned} \quad (14)$$

where $\rho_{RZA} = \mu\lambda_{RZA}\varepsilon_{RZA}$. The matrix-vector form of (14) can be rewritten as

$$w(n+1) = w(n) + \mu e(n)x(n) - \frac{\rho_{RZA}\text{sgn}(w(n))}{1+\varepsilon_{RZA}|w(n)|}, \quad (15)$$

where reweighted factor is set as $\varepsilon_{RZA} = 20$ [7] to exploit channel sparsity efficiently. In the second term of (15), please notice that channel coefficients $|w_i(n)|, i=0,1,\cdots,N-1$ are replaced by zeroes in high probability if under the threshold $1/\varepsilon_{RZA}$. Hence, one can find that SLMS-RZA can exploit

sparsity and mitigate noise interference simultaneously. Then the SLMS-RZA can be developed as

$$w(n+1) = w(n) + \mu \operatorname{sgn}(e(n))x(n) - \frac{\rho_{RZA}\operatorname{sgn}(w(n))}{1+\varepsilon_{RZA}|w(n)|}. \quad (16)$$

*C. Proposed algorithm: SLMS-RL1*

Beside the RZA-type algorithm, RL1 minimization for adaptive sparse channel estimation has a better performance than L1 minimization that is usually employed in compressive sensing [17]. It is due to the fact that a properly reweighted $\ell_1$ norm (RL1) approximates the $\ell_0$-norm, which actually needs to be minimized, better than $\ell_1$-norm. The cost function of LMS-RL1 was constructed as

$$G_{RL1}(w(n)) = (1/2)e^2(n) + \lambda_{RL1}\|f(n)w(n)\|_1, \quad (17)$$

where $\lambda_{RL1}$ is the weight associated with the penalty term and elements of the $1\times N$ row vector $f(n)$ are set to

$$[f(n)]_i = \frac{1}{\delta_{RL1}+|[w(n-1)]_i|}, \quad i=0,1,\cdots,N-1, \quad (18)$$

where $\delta_{RL1}$ being some positive number and hence $[f(n)]_i > 0$ for $i=0,1,...,N-1$. The update equation can be derived by differentiating (17) with respect to $w(n)$. Then, the resulting update equation of LMS-RL1 is

$$\begin{aligned}w(n+1) &= w(n) + \mu\frac{\partial G_{RL1}(w(n))}{\partial w(n)} \\ &= w(n) + \mu e(n)x(n) - \frac{\lambda_{RL1}\operatorname{sgn}(w(n))}{\delta_{RL1}+|w(n-1)|},\end{aligned} \quad (19)$$

Please notice that in Eq. (20), since $\operatorname{sgn}(f(n)) = \mathbf{1}_{1\times N}$, hence one can get $\operatorname{sgn}(f(n)w(n)) = \operatorname{sgn}(w(n))$. Note that although $w(n)$ changes in every stage of this sparsity-aware SLMS-RL1 algorithm, it does not depend on $w(n)$. Hence the cost function $G_{RL1}(n)$ is convex as well. According to (20), SLMS-RL1 is proposed as

$$w(n+1) = w(n) + \mu \operatorname{sgn}(e(n))x(n) - \frac{\lambda_{RL1}\operatorname{sgn}(w(n))}{\delta_{RL1}+|w(n-1)|}. \quad (20)$$

*D. Proposed algorithm: SLMS-LP*

Aforementioned three sparse SLMS algorithms are convex algorithms. Accurately, nonconvex sparse constraints, e.g., $\ell_p$-norm, can be also utilized to exploit channel sparsity. In [5], LMS-LP based adaptive sparse channel estimation method has been proposed to exploit channel sparsity efficiently. Similarly, the cost function of LMS-LP was constructed as

$$G_{LP}(w(n)) = (1/2)e^2(n) + \lambda_{LP}\|w(n)\|_p, \quad (21)$$

where $\lambda_{LP} > 0$ is a regularization parameter which balances the estimation error and channel sparsity. The corresponding update equation of LMS-LP is derived as

$$\begin{aligned}w(n+1) &= w(n) - \mu\frac{\partial G_{LP}(w(n))}{\partial w(n)} \\ &= w(n) + \mu e(n)x(n) - \frac{\rho_{RL1}\|w(n)\|_p^{1-p}\operatorname{sgn}(w(n))}{\varepsilon_{LP}+|w(n)|^{1-p}}\end{aligned} \quad (22)$$

where $\varepsilon_{LP} > 0$ denotes threshold parameter and $\rho_{LP} = \mu\lambda_{LP}$ is a parameter which depends on step-size and regularization parameter. According to the update equation of (23), SLMS-LP is proposed as

$$\begin{aligned}w(n+1) = &\,w(n) + \mu\operatorname{sgn}(e(n))x(n) \\ &- \frac{\rho_{RL1}\|w(n)\|_p^{1-p}\operatorname{sgn}(w(n))}{\varepsilon_{LP}+|w(n)|^{1-p}}.\end{aligned} \quad (23)$$

IV. NUMERICAL SIMULATIONS

In this section, the proposed sparse SLMS algorithms are evaluated in different scenarios: channel sparsity as well as impulsive noise environments. For achieving average performance, $M$=1000 independent Monte-Carlo runs are adopted. The simulation setup is configured according to typical broadband wireless communication system [3]. The signal bandwidth is 60MHz located at the central radio frequency of 2.1GHz. The maximum delay spread of $1.06\mu s$. Hence, the maximum length of channel vector $w$ is $N$=128 and its number of dominant taps is set as Sparsity $\in$ {4,8}. To validate the effectiveness of the proposed methods, average mean square error (MSE) standard is adopted. Channel estimators are evaluated by average MSE which is defined by

$$\operatorname{MSE}\{w(n)\} = 10\log_{10}(1/M)\sum_{m=1}^{M}\|w_m(n)-w\|_2^2/\|w\|_2^2 \quad (24)$$

where $w$ and $w_m(n)$ are the actual signal vector and reconstruction vector, respectively. The results are averaged over 1000 independent Monte-Carlo (MC) runs. Each dominant channel tap follows random Gaussian distribution as $\mathcal{CN}(0,\sigma_w^2)$ which is subject to $E\{\|w\|_2^2\}=1$ and their positions are randomly decided within the $w$. The received SNR is defined as $P_0/\sigma_n^2$, where $P_0$ is the received power of the pseudo-random noise (PN)-sequence for training signal. In addition, to achieve better steady-state estimation performance, reweighted factor of (S)LMS-RZA is set as $\varepsilon_{RZA} = 20$ [20]. Threshold parameter of (S)LMS-RL1 is set as $\delta_{RL1} = 0.05$ [5]. Detailed parameters for computer simulation are given in Tab. I.

In the first example, average MSE performances of the proposed methods are evaluated for *Sparsity*=8 in Figs. 1-2 under two SNR regimes (i.e., 10dB and 20dB) in the presence of impulsive noise ($\alpha = 1.2$). One can find the proposed sparse SLMS algorithms always achieve better performance than SLMS with respect to average MSE. In the case of SNR=10dB, the proposed sparse SLMS algorithm can achieve 3dB performance gain over standard SLMS as shown in Fig. 1. In the case of SNR=20dB, Fig. 2 shows that the proposed algorithms can still get 2dB performance gain. Hence, the effectiveness of the proposed algorithms is confirmed in the case of different SNR regimes.

TAB. I. SIMULATION PARAMETERS.

| Parameters | Values |
|---|---|
| Training signal | Pseudo-random Gaussian sequence |
| alpha-stable noise distribution | $\alpha \in \{1.2, 2.0\}, \beta = 0, \gamma = 1, \delta = 0$ |
| Channel length | $N = 128$ |
| No. of nonzero coefficients | $Sparsity \in \{4, 8\}$ |
| Distribution of nonzero coefficient | Random Gaussian $\mathcal{CN}(0,1)$ |
| Received SNR for channel estimation | $\{5dB, 10dB\}$ |
| Step-size | $\mu = 0.005$ |
| Regularization parameters for sparse penalties | $\lambda_{ZA} = 2 \times 10^{-4}$ $\lambda_{RZA} = 2 \times 10^{-3}$ $\lambda_{RL1} = 5 \times 10^{-5}$ $\lambda_{LP} = 5 \times 10^{-6}$ |
| Reweight factor of (S)LMS-RZA | $\varepsilon_{RZA} = 20$ |
| Threshold of the (S)LMS-RL1 | $\delta_{RL1} = 0.05$ |
| Threshold of the (S)LMS-LP | $\varepsilon_{LP} = 0.05$ |

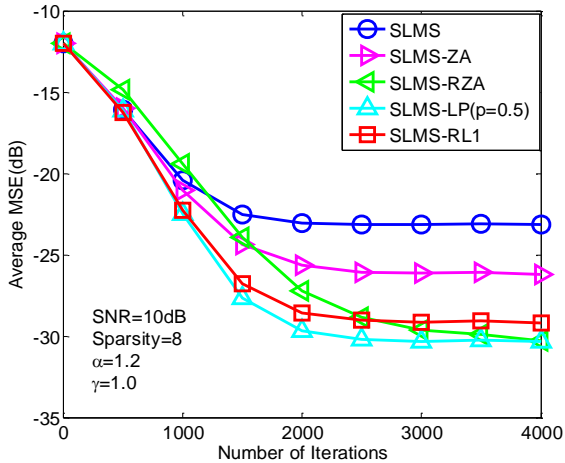

Fig. 1. Avergae MSE comparsions (SNR=10dB and $\alpha = 1.2$).

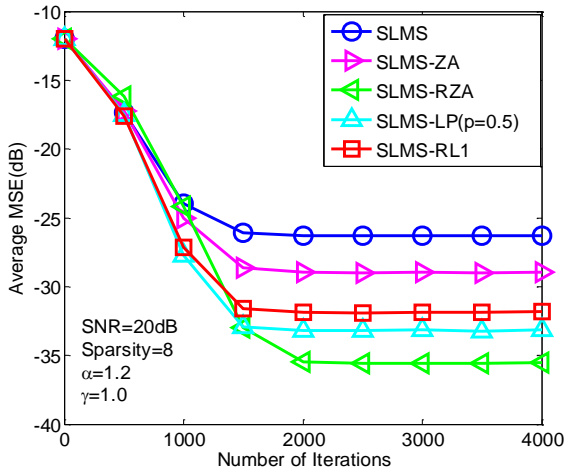

Fig. 2. Avergae MSE comparsions (SNR=20dB and $\alpha = 1.2$).

In the second example, the proposed methods are evaluated under Gaussian noise environment (i.e., $\alpha = 2.0$) in the case of SNR=10dB. In Fig. 3, MSE performance of these proposed methods are evaluated in *Sparsity*=4. The proposed SLMS algorithms are very close to sparse LMS ones. However, the proposed SLMS algorithms can achieve better than MSE performance than sparse LMS ones in the case of *Sparsity*=8. According to Figs. 3-4, one can deduce that the proposed SLMS algorithms can ensure stable for larger number of nonzero taps, e.g., *Sparsity*=16. Hence, the proposed algorithms are not only stable for impulsive noise interference but also for the large number of nonzero taps.

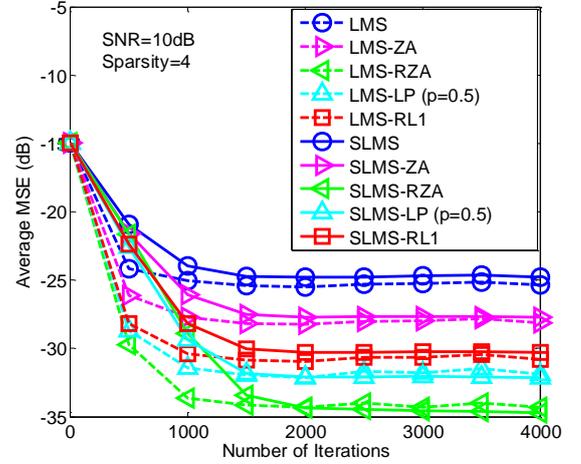

Fig. 3. Avergae MSE comparsions under the assumption of Gaussian noise.

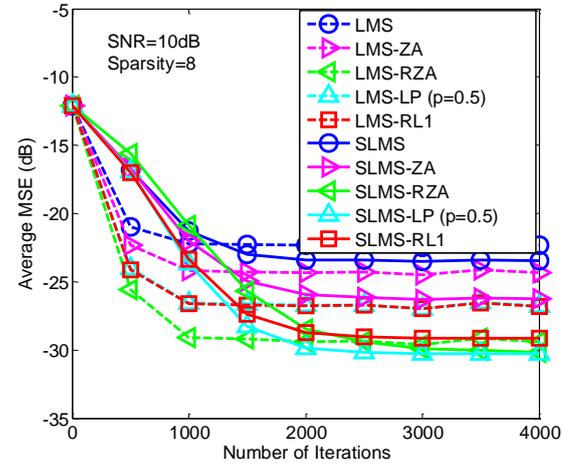

Fig. 4. Avergae MSE comparsions under the assumption of Gaussian noise.

## V. CONCLUSIONS AND FUTURE WORK

The family of alpha-stable distributions provides an accurate model of impulsive noise processes in communications channels and in fact, includes the Gaussian density as a special case. Based on this noise model, this paper proposed stable adaptive filtering algorithms by using different sparsity-inducing penalty functions. Computer simulation results are provided to verify the effectiveness of the proposed algorithms. In future work, we will test our proposed methods in different communications systems, such as underwater acoustic systems as well as power-line communication systems.